\documentclass[]{spie}  

\usepackage{amsmath,amsfonts,amssymb}
\usepackage{graphicx}
\usepackage[colorlinks=true, allcolors=blue]{hyperref}

\title{Astrometry with MCAO at Gemini and at ELTs}

\author[a]{Tobias K. Fritz}
\author[a]{Nitya Kallivayalil}
\author[b]{Eleazar R. Carrasco} 
\author[c]{Benoit Neichel}
\author[d]{Richard Davies}
\author[e]{Rachael Beaton}
\author[a]{Dylan Angell}
\author[a]{Sean Linden}
\author[a]{Paul Zivick}
\author[a]{Steve Majewski}
\author[a]{Guillermo Damke}
\author[f]{Mike Boylan-Kolchin}
\author[g]{Roeland van der Marel}
\author[h]{Tony Sohn}

\affil[a]{Department of Astronomy, University of Virginia,  3530 McCormick Road,Charlottesville, VA 22904, USA}
\affil[b]{Gemini South, Casilla 603
La Serena, Chile}
\affil[c]{LAM - Laboratoire d’Astrophysique de Marseille,
38, rue Frederic Joliot-Curie, 13388 Marseille, France}
\affil[d]{Max Planck Institut f{\"u}r Extraterrestrische Physik, Postfach 1312, D-85741, Garching, Germany}
\affil[e]{Observatories of the Carnegie Institution for Science,
813 Santa Barbara Street
Pasadena, CA 91101, USA}
\affil[f]{The University of Texas at Austin
Department of Astronomy,
2515 Speedway, Stop C1400
Austin, Texas 78712, USA}
\affil[g]{Space Telescope Science Institute,
3700 San Martin Drive,
Baltimore, MD 21218, USA}
\affil[h]{Department of Physics and Astronomy,
The Johns Hopkins University,
3400 N. Charles Street,
Baltimore, MD 21218, USA}

\authorinfo{Send correspondence to T. K. F. E-mail: tkf4w@virginia.edu}

\begin{document} 
\maketitle

\begin{abstract}
We present in this study a first analysis of the astrometric error budget of absolute astrometry relative to background galaxies using adaptive optics. We use for this analysis multi-conjugated adaptive optics (MCAO) images obtained with GeMS/GSAOI at Gemini South. We find that
it is possible to obtain 0.3 mas reference precision in a random field with 1 hour on source using faint background galaxies.
Systematic errors are correctable below that level, such that the overall error is approximately 0.4 mas. 
Because the reference sources are extended,  we find it necessary to correct for the dependency of the PSF centroid on the used aperture size, which would otherwise cause an important bias.
This effect needs also to be considered for Extremely Large Telescopes (ELTs). When this effect is corrected, ELTs have the potential to measure proper motions
of dwarfs galaxies around M31
with 10 km/s accuracy over a baseline of 5 years. 
\end{abstract}

\section{Introduction}
\label{sec:intro}  
Astrometry is useful for several astronomical purposes, with one primary application being the derivation of orbits and masses\cite{Kallivayalil13}. 
Until recently adaptive optics (AO) was mainly used for targets in which the interesting science is concentrated in a very small angular region like the Galactic Center\cite{Gillessen09}. This was mainly due to the technical constraints of the AO systems, which were single conjugated and can therefore correct the atmosphere only over a small field of view. The first laser-based multi conjugated AO system (GeMS/GSAOI\cite{Rigaut14,Carrasco12}) is now in operation at Gemini South. It has a much larger field of view than previous systems, and is thus useful for targets extended over a larger area of sky. We are currently using GeMS/GSAOI in a Long/Large Gemini program\footnote{http://www.gemini.edu/node/12335?q=node/12238\#Fritz} 
to determine absolute proper motions of objects (dwarf galaxies, globular clusters, stream stars) in the halo of the Milky Way. 
Since this is the first project of its kind it is important to characterize the astrometric errors well. We report here the current status of the error characterization and add some thoughts for ELTs.

\section{Fundamental limit for stars}
\label{sec:fun_limit}  
The fundamental limit of astrometry is set by the precision in the determination of the source center. In the case of point sources  observed with a diffraction-limited telescope with a circular aperture the equation is \cite{Lindegren78}: $\sigma_x=1/\pi\times \lambda/D\times 1/\mathrm{SNR}$.
 Therein, $\lambda/D$ stands approximately for the FWHM of the observations; thus when an observation is not diffraction-limited it can be replaced by the resolution of that observation. Usually this is the seeing; and this is one reason why
 seeing-limited observations have less astrometric precision than diffraction-limited observations. 
In addition, when an observation is background-limited, the SNR also depends on the resolution. The vast majority of sources are so faint that observations are background-limited in the near infrared. For a constant telescope size the signal is independent of the resolution, in contrast to the noise of the background which is proportional to the FWHM. Thus, for constant mirror size the following scaling relation is valid for astrometric precision: $\sigma_x\propto \mathrm{FWHM}^2$. 
 Therefore, fully diffraction-limited observations have much better astrometric precision than seeing-limited observations.
 It also follows that in the diffraction-limited case the precision is a strong function of the diameter: $\sigma_x\propto 1/D^3 $. 
  This means that the E-ELT (39 m) is about $7.6\times10^{6}$ times faster to reach a certain astrometric precision in Ks-band diffraction-limited mode than in seeing-limited (seeing$\approx600$ mas) mode. 
 Similarly, the E-ELT is about 12,000 times faster than Gemini when both are diffraction-limited.

\section{Gemini precision for stars}
\label{sec:real_prec} 
\begin{figure} [ht]
   \begin{center}
   \begin{tabular}{c} 
   \includegraphics[height=13cm, angle=-90]{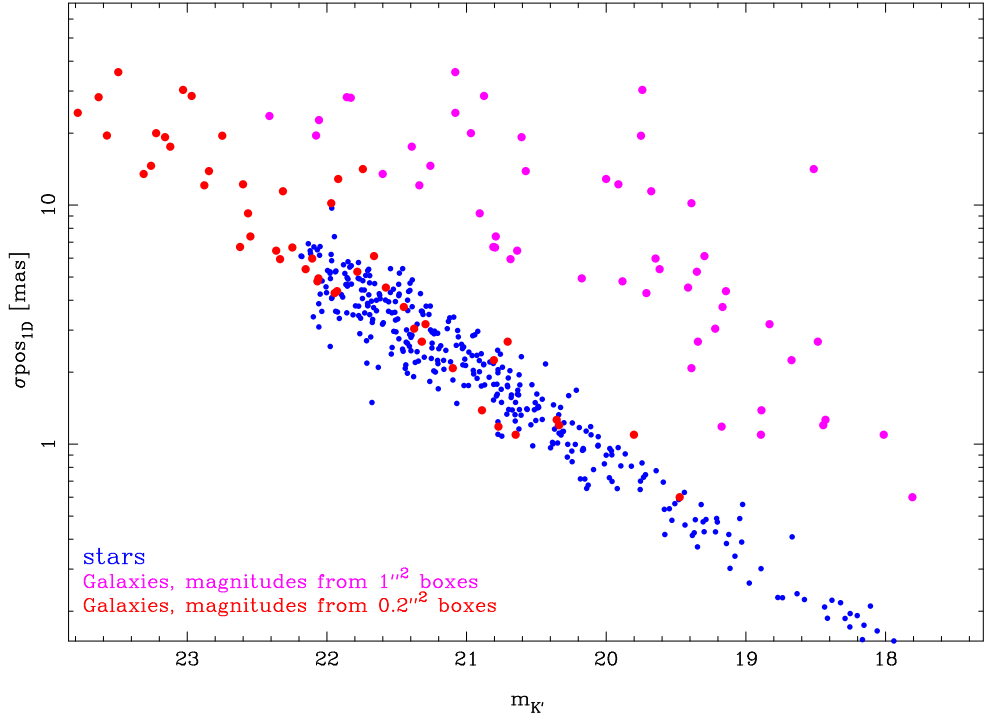}
   \end{tabular}
   \end{center}
   \caption[example] 
   { \label{fig:precision} 
Precision of stars and galaxies on a 3600 sec Gemini image. The magnitudes for stars are PSF magnitudes, while for galaxies we show  aperture magnitudes in two boxes of different sizes.
}
   \end{figure}

AO systems are not perfect in achieving a diffraction-limited correction, in particular systems which work with faint or off-axis stars. In practice only a fraction of the flux is concentrated in a PSF core which has a FWHM close to the diffraction limit. Usually the majority of the flux is in the halo, which has a FWHM similar to the seeing. This halo is essentially useless for stellar astrometry.
To get an empirical measure we use data centered on the globular cluster Pyxis, located in the outer halo of the Milky Way. We calculate the scatter of the positions of stars over 30 images, see Figure~\ref{fig:precision}. Our method involves fitting Gaussians to the images, which we first smooth with a Gaussian of 3 pixel FWHM. We also test other methods on these and simulated data, nearly all methods yield very similar precision for non-saturated stars. The standard positions of Sextractor\cite{Bertin96} are much worse, but other Sextractor options like `model', and `PSF-fit' position are as good as Gaussian fits and those from Starfinder\cite{Diolaiti00}. The precision of the windowed position from Sextractor 
is very similar; only for rather bright stars is it 20\% less precise.

We obtain a 1D precision of 0.14 mas for m$_{K'}=18$ stars after one hour of on-source integration. Because the noise is not caused by the stars themselves the position errors for other magnitudes follow $\sigma \propto 10^{0.4\, \mathrm{mag}}$. The Strehl ratio (SR) is on average about 17.5\% in the data. When we use the power law relation of Ref.~\citenum{Fritz09} for our calculation, an error of about 0.15 mas is predicted for our exposure time and noise properties. Thus, our data set and that of Ref.~\citenum{Fritz09} agree well. 
For one Carina field we have data in H and K'-band in roughly the same conditions.
We find that stars in that field have errors that are about 45\% smaller in H-band than in K'-band. While the SR is smaller in H-band, two other factors help to counter this: stars are intrinsically brighter in H-band and the background is fainter in the H-band.  A large part of the K'-band background is caused by the AO bench system Canopus\cite{Carrasco12}. A cold AO system like NFIRAOS\cite{Herriot14} would reduce the background. 

How much room is there for improvement? The relation of Ref.~\citenum{Fritz09} predicts an error of 0.056 mas at SR$=40$\% for our discussed case, while a fully diffraction-limited image with GeMS/GSAOI noise added would have an error of 0.009 mas.
Thus an improved SR has the potential to improve the astrometric precision. However, for many science cases the improvement can only be realized when the field of view remains large which is usually more difficult at high SR. 

   \section{Precision of galaxies}
\label{sec:gal_prec} 
  Galaxies are complex such that it is not possible to predict the precision a priori. We test the galaxy precision in several GSAOI fields by fitting them with Galfit \cite{Peng02} on a mosaic of our Pyxis data created using the THELI \cite{Schirmer13} package. 
  We use PSFs extracted from the image, the impact of the precise PSF on the galaxy precision (as opposed to any systematic effect) is anyway very low.  
  For the vast majority of the galaxies a model consisting of a single Sersic function fits the galaxies without systematic residuals. Only in a few cases does the fit improve clearly when a second component, Sersic or point source, is added to fit the central regions  better. We show the errors for the 52 galaxies
  in the Pyxis field 1 in Figure~\ref{fig:precision}. It shows that when we use a larger aperture to get nearly the full flux of the galaxies, the precision for galaxies is much worse than for stars for of the same magnitude. This is because galaxies are typically larger than stars which increases the FWHM and decreases SNR and thus reduces the positional accuracy. They also span a range of morphology i.e. where most of the flux is distributed, and due to the variation in the size also the scatter in the error is larger.
The scatter is smaller when only the central flux of these galaxies is used. 
  Besides very bright and large ($\approx$close) galaxies, the majority of galaxies which have the best position precision have a large Sersic index, and are thus early-type galaxies.
  
  The maximum possible precision in absolute astrometry is limited by the total precision of background galaxies. We obtain this total precision by error-weighted addition of the position error of all galaxies. We obtain values between 0.18 and 0.39 mas for the four fields currently being investigated. 
  For the one Carina field observed in both H and K'-band we find that the astrometric precision is identical in both bands within the uncertainties. This is different from the behavior of stars because galaxies are brighter in the K'-band because galaxies are redshifted.
  
To predict the precisions obtainable by ELTs for galaxies is very difficult, since the structure of galaxies at ELT resolution needs to be known. That is currently possible only for a few cases of lensed galaxies to a limited extent. Some kind of prediction of galaxy properties is thus necessary. Ref.~\citenum{Trippe08} used in their MICADO study for that purpose galaxies with many sub components. By using  cross correlation, 
they obtain using only a subset of the image, an error smaller than 5 $\mu$as. That strategy might not be the best due to systematics, see Section~\ref{sec:PSF}. Point sources should be less affected by these systematics, and by using star clusters Ref.~\citenum{Trippe08} obtain a precision of about 29 $\mu$as in one hour.
Another rough estimate can be obtained from the Gemini precision and a $D^x$ scaling law for the error. $x$ is -1 for fully extended sources at ELT resolution and -3 for sources which decompose fully into point sources in that case. Possibly, $x$ in the middle (x$=-2$) is a reasonable assumption. In the case of an instrument similar to 
 GSAOI mounted at the E-ELT, it would provide a precision of about 13 $\mu$as in one hour. 
 That estimate is pessimistic because it neglects that some faint galaxies are undetectable with Gemini but detectable with the E-ELT. 
Independent of what the achievable precision for galaxies is, it clear that a large field of view is desirable for absolute proper motions. The number of galaxies scales with area, and thus an instrument with a larger field of view ($A$) can reach the same precision faster. For the same telescope the necessary time scales as $ t\propto A^{-0.5}$.

   \section{Distortion}
\label{sec:dist}

Essentially all cameras have some distortion, which needs to be corrected.
Residual distortion is often the limiting error contribution\cite{Fritz09}.
We choose one data set, that of Pyxis 1, to test for residual distortion. 
As our roughly distortion-free reference we start with the THELI\cite{Schirmer13} mosaic. We measure stars in this image and in each of the single 30 Pyxis images with the same method as in Section~\ref{sec:real_prec}. We then find transformations from the single images to the mosaic treating each detector separately. We find that a quadratic transformation is sufficient. We iteratively improve these transformations, by giving the stars a weight according to $\sigma_\mathrm{pos}=\sqrt{(\mathrm{floor})^2+(c*10^{0.4\times(\mathrm{mag}-18)})^2}$. $\mathrm{floor}$ describes the systematic error floor probably caused by distortion.
We iteratively determine the parameters $\mathrm{floor}$ and $c$ by fitting the measured scatter.
 We also exclude some outliers. In the later iterations we choose as reference not the positions on the mosaic but the median of the transformed position of the single images. After 4 iterations the fit converges, the change in the fifth iteration is irrelevant.

For the distortion we test three different options for dealing with the time variability. In the first variant the distortion is constant. In the second it is constant for images with less than 0.8" relative offset. For these offsets the astrometric loop is closed. In our observing sequences we observe usually six consecutive images with such small offsets and make then a bigger offset of about 5". In the third variant we calculate the distortion field separately for each image. Our definition of the distortion is that it consists only of the non-linear terms. We calculate the linear terms and the offset for each image in all three variants. As a measure of the distortion error we use the median scatter for the 18 stars with $m_{K'}<16.4$. These stars are so bright that they are not limited by their SNR.
For a single image, that scatter decreases from 0.40 mas to 0.29 mas to finally 0.24 mas in x and in y from 0.26 mas to 0.24 mas and 0.22 mas when allowing for more time variability. It seems that the impact of the astrometric loop is not so large in our case, implying that the system is stable.
For our final positions the scatter is even smaller because we add positions from up to 
30 images together to the final measurement of one epoch. 
Assuming that the residuals are random we get a distortion error of about 0.04 mas. Since we use a quadratic transformation, about 7 well-measured sources are necessary to achieve that precision.
The discussion above assumes that the mosaic is distortion-free, and we have not yet tested to what extent that is the case. This, however, provides a first characterization of the expected size of the distortion. 

The distortion stability was measured over less than 3 hours. We have not yet suitable data to measure it over longer time scales.
For the large program, variability over at least a 2 year time scale will be relevant. Current research indicates residuals of about 0.4 mas over about one month\cite{Neichel14}.

   \section{Differential chromatic refraction}
\label{sec:dcr} 

The atmosphere refracts the light of the targets significantly. The biggest part of that refraction is constant, it is thus irrelevant for our aim of relative astrometry. There is some differential refraction with dependence on the zenith angle but the applied linear transformations correct for it\cite{Fritz09}. Even in the case of our large field of view, which can have up to 120" diameter, the non-linear effect is only 7 $\mu$as even at a large zenith angle of $45^\circ$.

\begin{figure} [ht]
   \begin{center}
   \begin{tabular}{c} 
   \includegraphics[height=8.4cm, angle=-90]{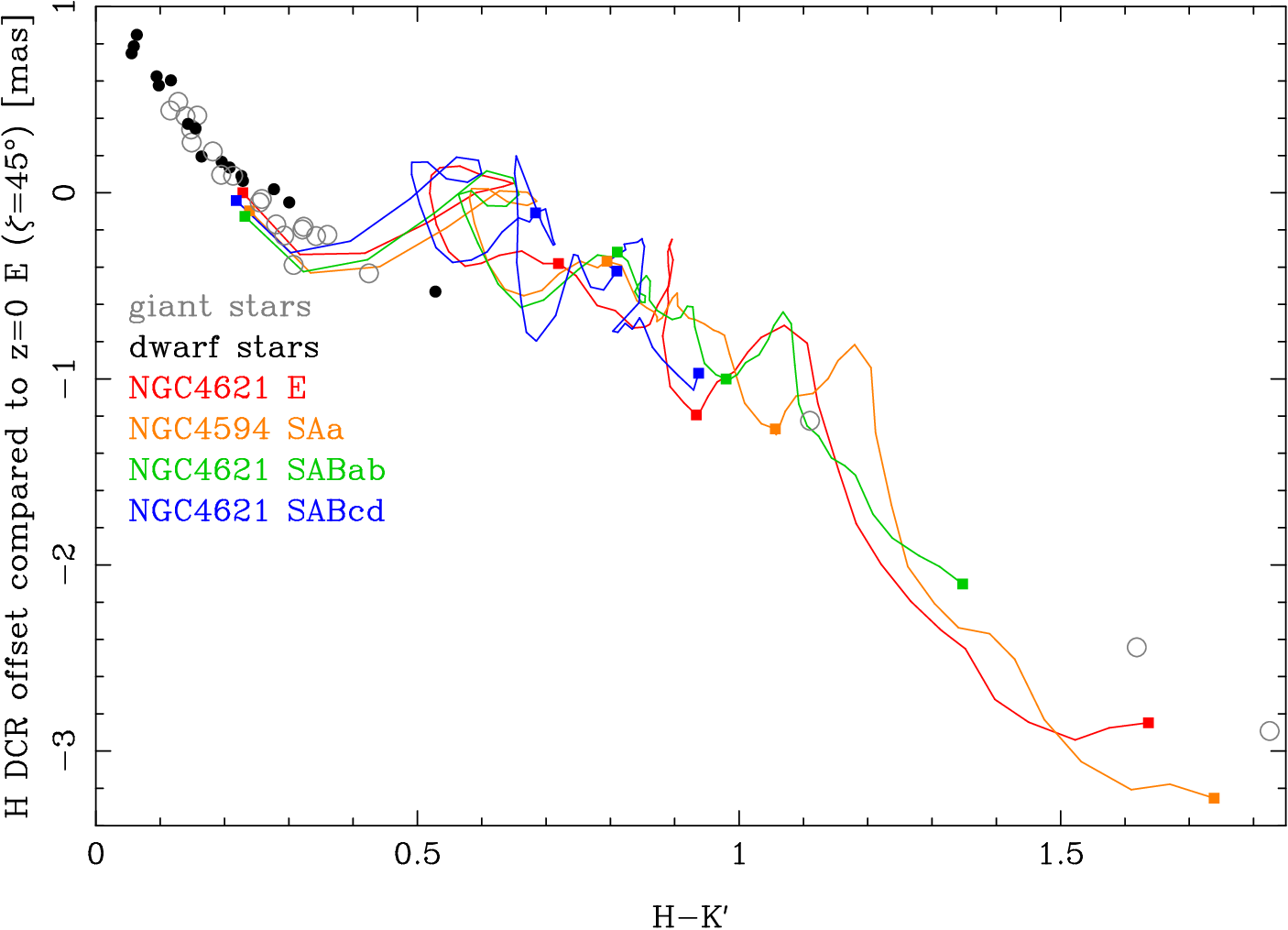}
   \includegraphics[height=8.4cm, angle=-90]{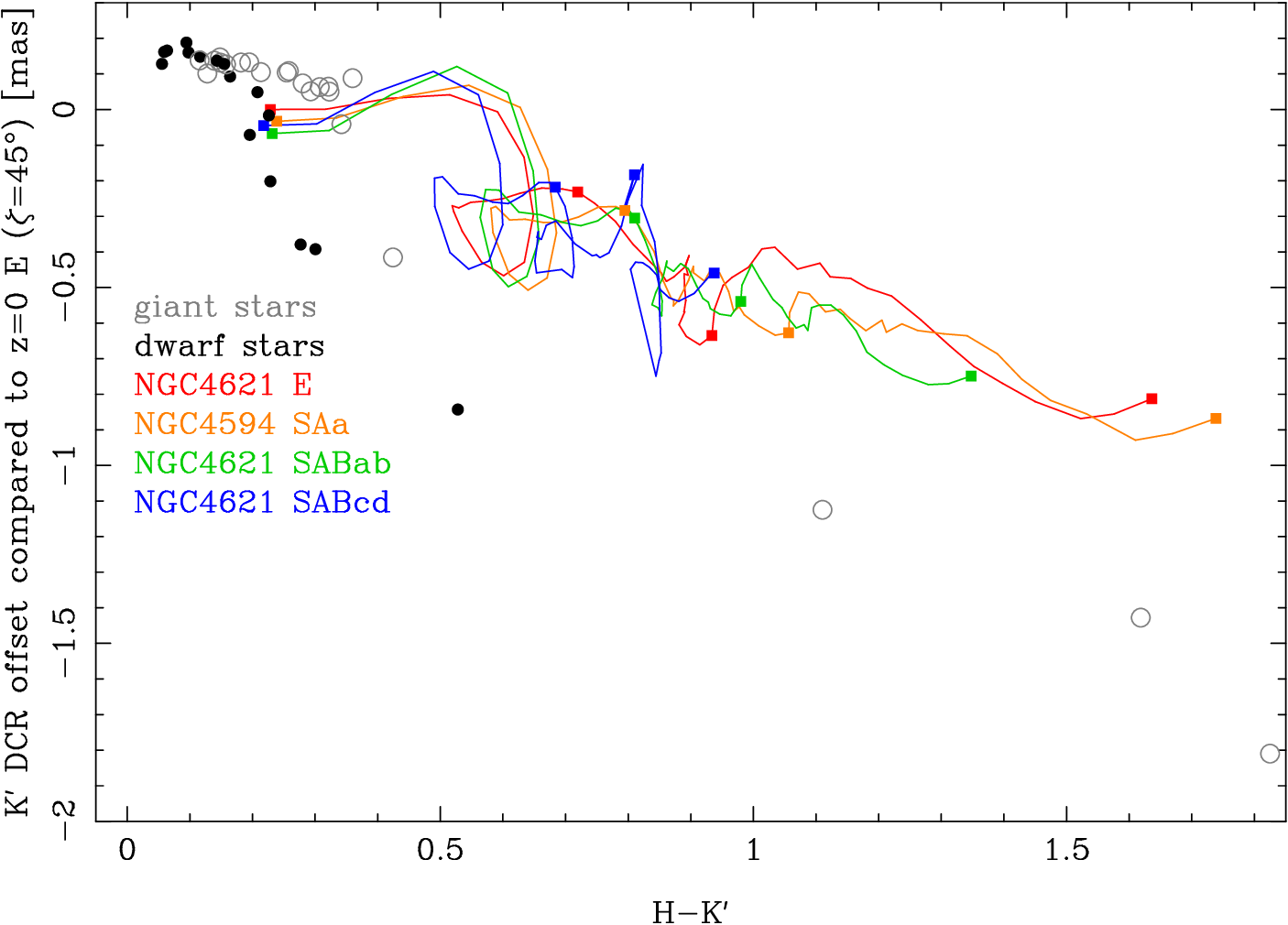}  
   \end{tabular}
   \end{center}
   \caption[example] 
   { \label{fig:dcr} 
Differential chromatic refraction in the H-band (left) and K'-band (right). The lines represent galaxies redshifted in steps of $\delta\mathrm{z}=0.1$ from z$=0$ to z$=3$.  The colored boxes on the lines mark from the left to the right the integer redshifts between 0 and 3.
}
   \end{figure}

However, there is additional differential chromatic refraction (DCR) because the atmosphere refracts light of different colors with different strengths. Since the celestial objects do not have the same colors this effect needs to be considered. The wavelength dependence in the optical is rather large and DCR can lead to position shifts of about 4 mas \cite{Fritz15}. According to Ref.~\citenum{Edlen53} the refraction index as a function of wavelength in the H and K-band can be derived from $n_1\times10^7=2029.94-2.87\lambda_2+2.16\lambda_2^2-1.44\lambda_2^3+0.92\lambda_2^4$ where $n_1=n-1$ and $\lambda_2=\lambda[\mu\mathrm{m}]-2$. Similar to Ref.~\citenum{Fritz15} we use spectra of stars and galaxies to calculate the DCR. We use star spectra from Ref.~\citenum{Rayner09}. As in Ref.~\citenum{Fritz15} we calculate synthetic colors and differential refraction terms from those spectra.
The sample does not include spectral types of A-type or earlier stars but they are very rare in our target fields. In addition the effective wavelength and thus the DCR does not change much among relatively hot stars because they all  follow rather well the Rayleigh-Jeans law in H and K'-band.
For galaxies it is difficult to get deep enough spectra of galaxies similar to our reference galaxies. 
Instead we start with galaxies with very low redshift from Ref.~\citenum{Brown14}, precisely we use NGC4621 (E), NGC4594 (SAa), NGC4569 (SABab) and NGC4559 (SABcd). These spectra are real spectra in the optical but are model spectra in the near infrared.
We redshift these spectra to model galaxies at higher redshift and calculate the same properties as for stars.
The results are shown in Figure~\ref{fig:dcr}.

The DCR shifts are smaller in the K'-band, because the following three effects all reduce the chromatic effects in the K'-band: the differential refraction decreases with wavelength, the smaller relative width of the K'-band compared to the H-band, and the spectra of most objects show less wavelength dependency at redder wavelengths. The latter effect is especially important for stars, since most stars are already in the Rayleigh-Jeans tail in the K'. (The exception are the red M-dwarfs and partly also M-giants, but halo targets should not contain any observable M-dwarfs at Gemini depth and only very few M-giants.) 
Galaxies show a larger range of colors in general, but in our case the majority have about $H-K'\approx0.8$. 
Relative to our typical target stars this leads to a DCR difference for the reference galaxies of about 1 mas in H-band and about 0.5 mas in K' at a zenith angle of 45$^\circ$. Most of our observations have a far smaller zenith angle and are observed in K'-band, reducing the typical effect of DCR to 0.1 - 0.2 mas even before correction.
Since color and offset are correlated a partial DCR correction is possible with H-K' colors. 
Further, since most of our objects are observed over a range of hour angles and airmasses and DCR affects only the position perpendicular to the horizon, it is possible to correct for DCR using only the images in the main band. 
After these corrections the residual DCR should be really small, probably less than 0.1 mas.

\section{PSF effects}
\label{sec:PSF} 

In this section, we summarize the impact of the PSF on the determination of the position of the center of a star or galaxy. 
For relative astrometry, we do not care about how the center is defined. It is only necessary that in each image the measured position of the object has a consistent offset compared to the ``true'' position.
Multi-year AO proper motion studies \cite{Trippe08,Gillessen09,Fritz14} have shown that the impact of PSF effects on consistent centering is smaller than 0.5 mas for stars even when using simple methods like Gaussian fitting, in the regime where the detector is well-sampled. For MCAO there are fewer studies, but Ref.~\citenum{Meyer11} showed using MAD data that the scatter is 1 mas. This is roughly consistent with Ref.~\citenum{Fritz14}. In both cases the scatter is likely caused mostly by distortion.
Our distortion analysis (Section~\ref{sec:dist}) shows 
that at least within one epoch the error floor for our GeMS/GSAOI data is about 0.3 mas. 

Typically, an implicit assumption is that the PSF effects on galaxy centers are the same as on star centers. But since galaxies are extended, while stars are point sources, it is possible that the PSF influences the centers in different ways. 
We determine whether there is an offset between star and galaxy centers in the following way. We derive galaxy shapes from images by fitting them with single a Sersic profile using Galfit. We use two different data sets, firstly the GSAOI observations of Pyxis field 1 and secondly, simulated observations. The simulated data starts with H160W observations of the UKIRT Ultra Deep Survey\cite{Lawrence07} (UDS). The H160W data is then deconvoled 
with the HST PSF using the MISTRAL\cite{Fusco03} method,
we then
subtract the stars from it, and add the stars from the Pyxis field. 
The image is then convolved with a GeMS/GSAOI PSF grid obtained with a a full end-to-end Monte Carlo simulation derived from YAO\footnote{\url{http://frigaut.github.io/yao/index.html}}.
Finally, the  noise, with properties consistent with our GSAOI data, is added to the image. 
The results using both the simulated image and the Pyxis image are, in essence, similar. We report here mainly the results from Pyxis, since that is based on sky data. 
  
We use the measured Sersic parameters to generate high SNR galaxies whose centers we know. We then convolve them with PSFs. These PSFs are usually simply isolated bright stars since 
crowding effects from neighboring sources is not a relevant issue for our Pyxis data.  
For each PSF we generate a star as control reference simple by using the PSF.
We then measure the object centers in two different ways. As the first option, we use Galfit, which takes a PSF model as input, and which is then convolved with the image. For galaxies the used model in Sersics, for the control star it is the PSF.
As the second option, we fit a Gaussian 
directly to the images without using a PSF. 
We fit the reference star in the same way.
For realistic PSF knowledge the first method is not better than the second.  
Below we report results from the second method.
In our test here,
the positions put into Galfit are the truth, and measure the offset between these positions and those obtained with a simple Gaussian. If that offset would be the same for stars and galaxies there would be no object class specific PSF effect.
However, stars and galaxies have different offsets, $\delta \mathrm{offset}_\mathrm{type}=\mathrm{offset}_\mathrm{gal}-\mathrm{offset}_\mathrm{star}=\mathrm{gal\,pos}_\mathrm{measured}-\mathrm{gal\,pos}_\mathrm{input}-(\mathrm{star\,pos}_\mathrm{measured}-\mathrm{star\,pos}_\mathrm{input})\neq0$

Why is that the case? The reason is that our PSFs are
not symmetric. That has the consequence that the position of the PSF center depends on its definition. The center of the core, which can be obtained approximately by fitting a Gaussian to the central $6 \times 6$ pixels is usually not identical with the centroid of the full PSF. The former is very close to what Galfit and other fitting codes derive as the centers of stars, because the SNR in the wings is very low. 
In contrast, for objects much larger than the PSF, 
one finds the center using a centroid, and this is affected by the wings of the PSF which are in regions with high SNR.
 Our reference galaxies are not so large, they are somewhere in between.

      \begin{figure} [ht]
   \begin{center}
   \begin{tabular}{c} 
   \includegraphics[height=8.4cm, angle=-90]{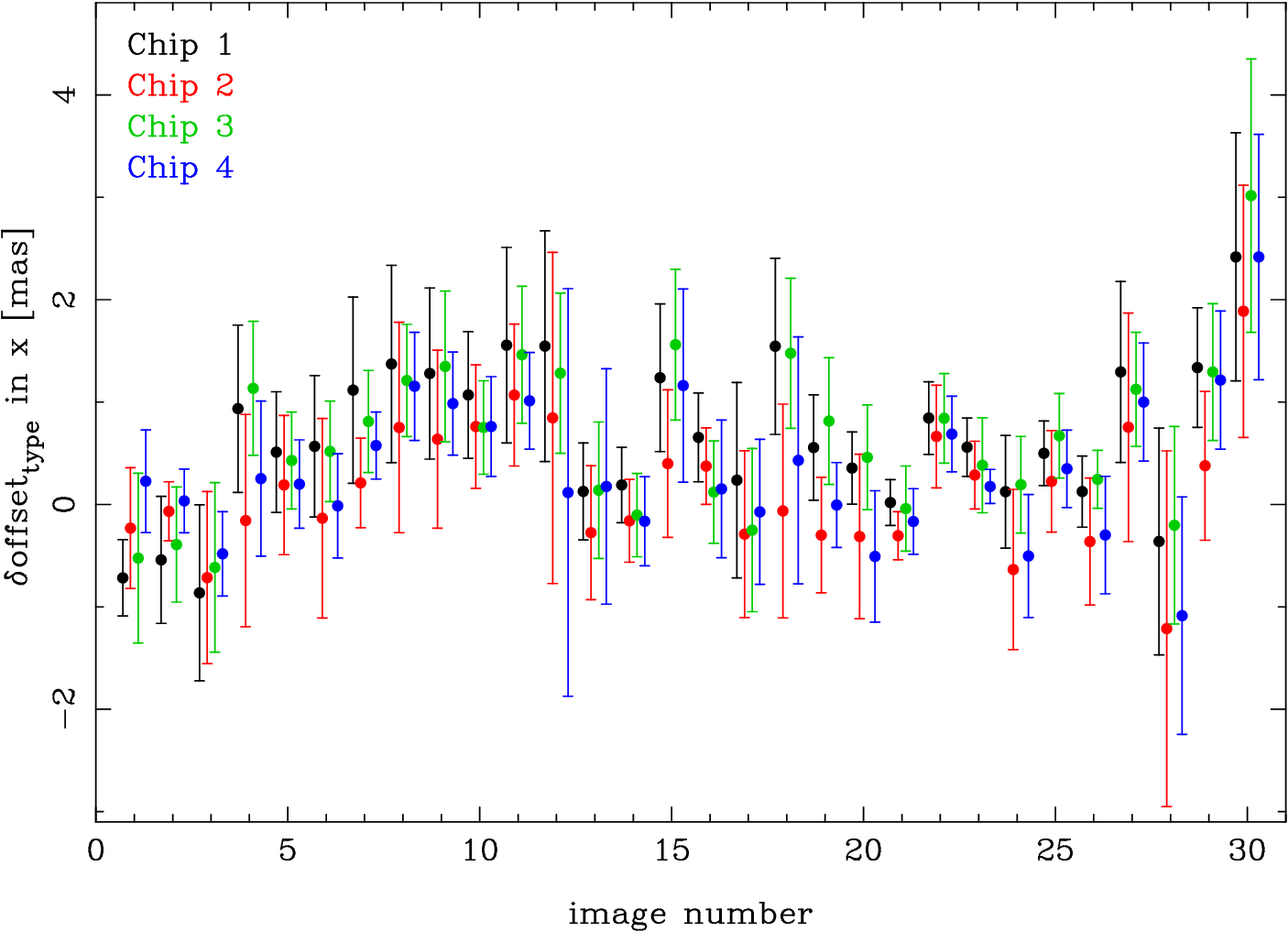}
   \includegraphics[height=8.4cm, angle=-90]{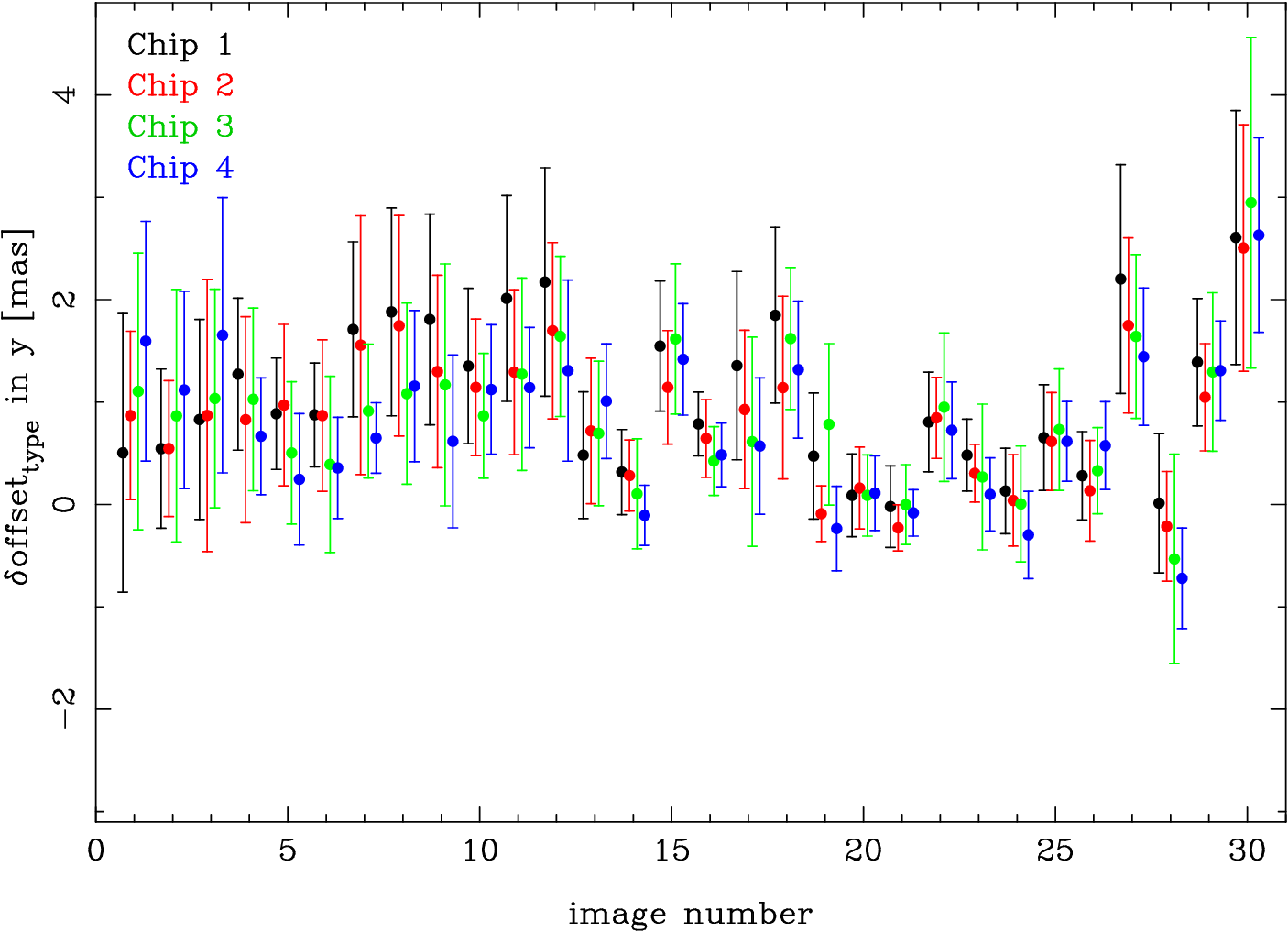}  
   \end{tabular}
   \end{center}
   \caption[example] 
   { \label{fig:psf_effect} 
Position offset of a typical galaxy relative to a star in x and y for our 30 images. (The points are slightly shifted in x to reduce crowding.) The shown values are obtained by interpolation. For each image, the shift is measured at the same pixel positions, and the median and the scatter is plotted. 
}
   \end{figure}

   The 
$\delta \mathrm{offset}_\mathrm{type}$  
depends on the PSF, but in a complicated way. The correlation with SR is very weak with maybe the exception of SR$<12$\%. We show in Figure~\ref{fig:psf_effect} the typical $\delta \mathrm{offset}_\mathrm{type}$  per image. For that we use the 15 galaxies with largest position precision and calculate their median $\delta \mathrm{offset}_\mathrm{type}$.  
To avoid the effect of the differential sampling of the image by the PSF stars, we interpolate between the measured values and show interpolations always at the same pixel
              positions. The median $\delta \mathrm{offset}_\mathrm{type}$   is 0.5 mas in x and 0.9 mas in y. There is a scatter of about 1.5 mas but over our 30 images, the final error reduces to the level of 0.3 mas which is less important than that since the median $\delta \mathrm{offset}_\mathrm{type}$   is not zero and it causes a bias.
The $\delta \mathrm{offset}_\mathrm{type}$   varies greatly from galaxy to galaxy. 
Some galaxies have a median $\delta \mathrm{offset}_\mathrm{type}$ bias  of up to about 3 mas. Usually these galaxies with a large $\delta \mathrm{offset}_\mathrm{type}$ bias  are relatively large and have poor position precision.
For the same PSF essentially all galaxies have the same direction for the offset, the direction of the PSF asymmetry. Thus, this effect does not average out over many galaxies.
It  turns out that the  average $\delta \mathrm{offset}_\mathrm{type}$ bias   is larger in our analysis of the UDS data than for our Pyxis data (2.4 versus 0.9 mas). 
That could be caused by cosmic variance, color effects between H160W and K', or the difference in resolution between GSAOI and HST.
 In any case even the $\delta \mathrm{offset}_\mathrm{type}$  bias derived from the Pyxis data is too large to be ignored. It is necessary to correct for it.

We now test whether it is possible to use interpolation of the $\delta \mathrm{offset}_\mathrm{type}$ of the stars to predict the $\delta \mathrm{offset}_\mathrm{type}$  at other locations of the image to a sufficient accuracy. 
For interpolation we use the Kringe algorithm. We use Jackknife (i.e., excluding always one PSF per image) to test the goodness of interpolation. We find that spherical Kringe interpolation works best. The usual deviation of  the Jackknife prediction from the measurement 
is about 0.55 mas. Since this is random it does not cause a bias and as such averages down to about 0.1 mas in the final position. Even in the case that this is somewhat underestimated its impact is small on the galaxies in the Pyxis data. However, that is likely not the case for all our Large program targets, since other targets have fewer stars bright enough to measure the offset well. For very sparse fields PSF reconstruction\cite{Villecroze12,Jolissaint14} might be a way forward in the future, when it works well enough with MCAO. 
Maybe it is also possible to improve the AO system such that the delivered PSF is more symmetric. Besides it might be advisably to give point sources more weight in the reference targets.
That is difficult for GeMS at Gemini, but ELTs might be able to use super star clusters as references \cite{Trippe08}.

   \section{Conclusions}
\label{sec:con}

In this work we analyze the error contribution to GeMS MCAO absolute astrometry mainly in one example case. We characterize the following important error contributions:

\begin{itemize}
 \item Random errors of stars are not relevant for Pyxis field 1. However, a few of the target fields in our program have fewer stars. Therefore the added precision of all stars will contribute more significantly to the error budget.
 \item The random errors of galaxies are larger. Although cosmic variance causes some scatter, the astrometric reference frame error integrated over all galaxies in one GSAOI field is usually about 0.3 mas in one hour on-source integration. Longer integration decreases this error. 
 \item Within one observation, star positions can be measured to 0.04 mas precision relatively. 
There is distortion variability on longer time scales, which is currently not corrected for at the same level. Since for our absolute motions several target stars and reference targets are averaged, the impact of residuals of 0.4 mas as in Ref.~\citenum{Neichel14} can be averaged down to about 0.2 mas. 
 \item Differential chromatic refraction can be corrected to less than 0.1 mas using color information and the images themselves. 
 \item PSF shape uncertainties have an impact of about 0.1 mas in our example case. The impact is probably larger for more sparse fields.
 \end{itemize} 
 Summarizing we expect an overall one epoch error of 0.4 mas for the case discussed, Pyxis field 1.
   
   \begin{figure} [ht]
   \begin{center}
   \begin{tabular}{c} 
   \includegraphics[height=13cm, angle=-90]{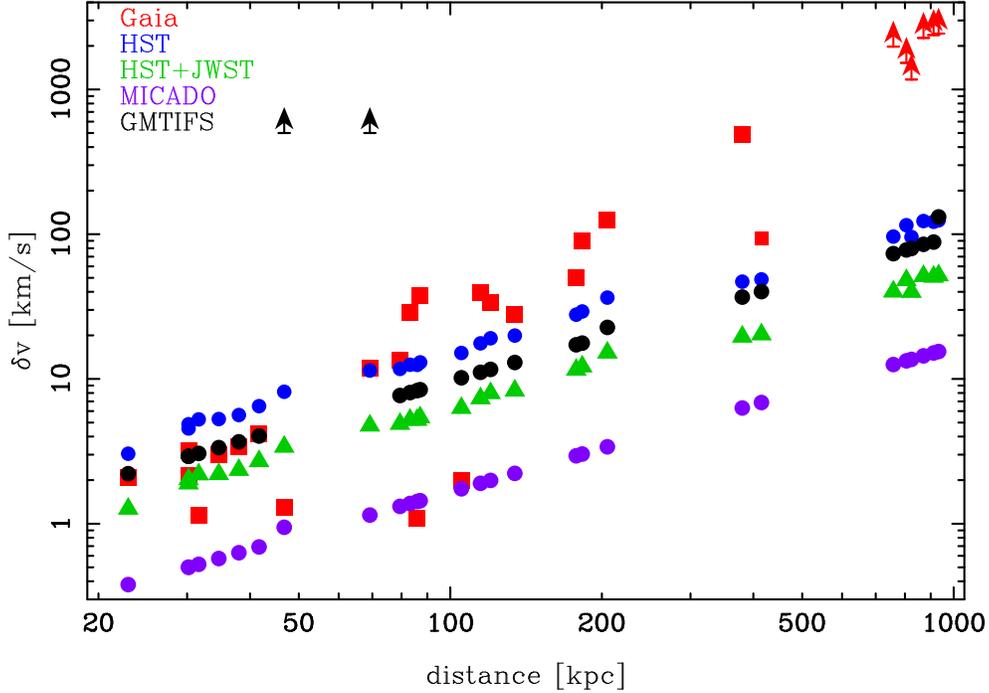} 
   \end{tabular}
   \end{center}
   \caption[example] 
   { \label{Fig:abs-mot} 
Absolute motion errors for some dwarf galaxies in the local group. The Gaia results assume the final data release, HST is calculated for a 5 year baseline, HST-JWST for a 12 year baseline and the ELT measurements assume a 5 year baseline with each 4 hours on source. Each galaxy is shown once for each instrument, the large scatter of Gaia is because Gaia is star limited while the other cases are mainly reference frame limited in precision.
}
   \end{figure} 
   
ELT instruments account for some of these systematic error sources already using specific hardware within the instruments
The PSF characterization could be the exception.  
Development of new approaches to characterize the PSF may be necessary to reduce its impact on the error budget, since this error source probably does not scale down with mirror size. The large size of ELTs reduce statistical errors compared to 8 m class telescopes.
 We here show as an example case the power of ELTs in measuring absolute proper motions  throughout the local group, see Figure~\ref{Fig:abs-mot}\footnote{For reference system errors we use the relation from Section~\ref{sec:gal_prec} for ELTs, empirical data on galaxy precision\cite{Sohn13} for HST (and JWST) and assume that this irrelevant for Gaia.  We also account for the expectable number of stars and their errors, it is rather unimportant for ELTs, for which we set its contribution to irrelevant when 6 stars with sufficient SNR are detected, and HST but has a large impact on Gaia.}. It is visible that while for some dwarf galaxies Gaia is very competitive, this is not the case for most, because they do not contain enough stars which are well measurable by Gaia. It is also clear that a large field of view, which MICADO\cite{Davies10} has in contrast to GMTIFS\cite{McGregor12}, is very useful for measuring absolute proper motions.

\acknowledgments 

This work was supported in part by NSF grant ID number 1455260.\\
Based on observations obtained at the Gemini Observatory, which is operated by the 
Association of Universities for Research in Astronomy, Inc., under a cooperative agreement 
with the NSF on behalf of the Gemini partnership: the National Science Foundation 
(United States), the National Research Council (Canada), CONICYT (Chile), the Australian 
Research Council (Australia), Minist\'{e}rio da Ci\^{e}ncia, Tecnologia e Inova\c{c}\~{a}o 
(Brazil) and Ministerio de Ciencia, Tecnolog\'{i}a e Innovaci\'{o}n Productiva (Argentina).
\bibliography{report2} 
\bibliographystyle{spiebib} 

\end{document}